\documentclass[%
 aip,
 jap,%
 amsmath,amssymb,
 reprint,%
]{revtex4-1}

\usepackage{graphicx}
\usepackage{dcolumn}
\usepackage{bm}

\begin{document}

\title[Brownian Simulations]{Simulations of magnetic nanoparticle Brownian motion}

\author{Daniel B. Reeves}
 \affiliation{Department of physics and astronomy, Dartmouth College, Hanover, NH 03755, USA}
 \email{dbr@Dartmouth.edu.}
 
\author{John B. Weaver}%
 \affiliation{Department of physics and astronomy, Dartmouth College, Hanover, NH 03755, USA}
 \altaffiliation[]{Radiology Department, Geisel School of Medicine}

\begin{abstract}
Magnetic nanoparticles are useful in many medical applications because they interact with biology on a cellular level thus allowing microenvironmental investigation. An enhanced understanding of the dynamics of magnetic particles may lead to advances in imaging directly in magnetic particle imaging (MPI) or through enhanced MRI contrast and is essential for nanoparticle sensing as in magnetic spectroscopy of Brownian motion (MSB). Moreover, therapeutic techniques like hyperthermia require information about particle dynamics for effective, safe, and reliable use in the clinic. To that end, we have developed and validated a stochastic dynamical model of rotating Brownian nanoparticles from a Langevin equation approach. With no field, the relaxation time toward equilibrium matches Einstein's model of Brownian motion. In a static field, the equilibrium magnetization agrees with the Langevin function. For high frequency or low amplitude driving fields, behavior characteristic of the linearized Debye approximation is reproduced. In a higher field regime where magnetic saturation occurs, the magnetization and its harmonics compare well with the effective field model. On another level, the model has been benchmarked against experimental results, successfully demonstrating that harmonics of the magnetization carry enough information to infer environmental parameters like viscosity and temperature.
\end{abstract}

\maketitle

\section{Introduction to Brownian nanoparticles}

There is a wide range of possible uses for magnetic nanoparticles (MNPs) in medical applications. Novel modalities image particles themselves as in magnetic particle imaging (MPI) \cite{mpi} and particles are also used as contrast agents in conventional magnetic resonance imaging \cite{pank}. Magnetic spectroscopy of Brownian motion (MSB) uses particle dynamics to gather information about the microscopic environments the particles inhabit \cite{weav}. A therapeutic method called hyperthermia damages unwanted cells (e.g in cancer treatment) by depositing energy from particles that are unable to rotate in time with oscillating fields due to viscous drag or anisotropic energy barriers  \cite{hypertherm}. In each of these techniques, it is necessary to understand the rotational dynamics of the particles in magnetic fields. Furthermore, simulation studies can be used by practitioners to choose ideal nanoparticle conditions and satisfy their specific needs \cite{pank}. 

In this work we examine Brownian rotation where particles physically rotate and experience viscous drag when placed in an oscillating field. The physics of Brownian rotation is governed by the average time it takes a particle to relax to an equilibrium value given its surroundings. The Brownian time constant $\tau_B$ is based accordingly on the viscosity $\eta$ and temperature $T$ of the suspending fluid, the hydrodynamic particle volume $V$, and Boltzmann's constant $k_B$ \cite{WFB,Ein}:
\begin{equation}\tau_B = \frac{3\eta V}{k_B T}. \label{TB}\end{equation} 
Brownian rotation is the dominant physical process when the particles are thermally blocked. This means that the energy barrier for internal magnetic moment relaxation is greater than the thermal energy of the system. If the particle is not thermally blocked, the electronic spins within single magnetic domains can rotate in unison, switching the magnetization of the particle. This spin rearrangement is referred to as N\'{e}el relaxation, and as described in Ref.~\cite{WFB} has a time constant of $\tau_N = \tau_0\exp(KV_c/k_BT)$ where $K$ is the magnetic anisotropy constant, $V_c$ the particle core volume, and $\tau_0$ the N\'{e}el attempt time characteristic of the material. Therefore, if a measurement is taken on a scale shorter than the relaxation time, no N\'{e}el spin flips are expected on average. 

It is important to note also that both relaxation processes occur without an applied field as the probabilities of random relaxation increase with heightened thermal activity. In practice, because the particle volume is a defining element of the blocking energy, the two regimes are often separated by the size of the particles. Therefore, when we consider larger 100nm diameter particles like those used in magnetic spectroscopy we are confident the dominant mechanism of relaxation is Brownian.

\section{Stochastic modeling from a balance of torques}

If we consider a dilute sample of MNPs we can imagine each as a separate isotropic dipole with a single magnetization direction determined by an internal crystalline properties. In a vacuum, the magnetic torque on such a dipole is expressed as the cross product between its magnetic moment and the field, $\mathcal{T}=\mathbf{m} \times \mathbf{B}$. The magnetic moment can be found from the material saturation magnetization multiplied by the magnetic core volume $M_sV_{core}$. Nanoparticles are commonly dispersed in a fluid so we must account for viscous drag during rotation. To first approximation, this torque is proportional to the angular velocity of the particle with magnitude given by the Stokes-Einstein relation for small Reynold's number particles \cite{Ein}. Written, the torque is $\mathcal{T} = -6\eta V \mathbf{\Omega}$, with $\mathbf{\Omega}$ the angular velocity, $V$ the hydrodynamic volume of the particle, and $\eta$ the fluid's viscosity. 

At the nanoscale we must also consider the powerful influence of thermal effects caused by random collisions of nanoparticles with the minute molecules of the fluid. If we assume that the time between such interactions is much shorter than the reaction time of the particles, we are free to consider the statistical fluctuations Markovian -- uncorrelated spatially or temporally. Accordingly we also implement a white noise force $\mathcal{N}$ in our model \cite{Lemons}. White noise is characterized theoretically by a flat power spectrum in frequency so that in Fourier space the behavior is delta autocorrelated in time and space with zero mean value. Explicitly, \begin{equation} \langle\mathcal{N}(t)\rangle=0 \hspace{.5cm} \langle\mathcal{N}_i(t)\mathcal{N}_j(t')\rangle=2D\delta_{ij}\delta(t-t').\end{equation}

Thus the expectation value of our random force at any time is zero, and no previous force affects the subsequent steps. The Einstein-Smoluchowski diffusion constant $D$ for spherical rotations depends on both the thermal energy as well as the viscous drag $D=6k_BT\eta V$. This may be determined from the fluctuation dissipation theorem or the associated Fokker-Planck equation for the dynamics \cite{garpal}. We now have the complete balance of torques 
\begin{equation} \mathcal{T} = \mathbf{m} \times \mathbf{B} -6\eta V \mathbf{\Omega}+\sqrt{2D}\mathcal{N}.\end{equation}

The acceleration term proportional to the moment of inertia can be ignored because we have already specified a low Reynolds number for the MNPs.  This means that the frictional drag forces are sufficiently intense as to prevent inertial rotation. With this assumption we can simplify considerably to a first order differential equation
\begin{equation} \mathbf{\Omega} = \frac{1}{6\eta V}\left( \mathbf{m} \times \mathbf{B} +  \sqrt{2D}\mathcal{N} \right). \label{tbal} \end{equation}

We can further clarify the expression by noting that the magnetic moment vector will experience a tangential rate change determined by the perpendicular component of the angular velocity of the moment viz. $d\mathbf{m}/dt = \mathbf{\Omega \times \mathbf{m}}$ \cite{TM}. Using this in Eq.~\ref{tbal} we obtain the full stochastic Langevin equation governing rotational dynamics of a magnetic Brownian particle suspended in fluid and placed in a magnetic field \cite{Raib,GWz}:
\begin{equation} \frac{d\mathbf{m}}{d t}= \frac{1}{6\eta V}\left( \mathbf{m} \times \mathbf{B} +  \sqrt{2D}\mathcal{N}\right) \times \mathbf{m}. \label{sde}\end{equation}

We chose to make no further assumptions, instead resorting to stochastic numerical analysis. The construction of the stochastic model is described below. Once prepared, we embarked on a series of simulations in order to benchmark against theories in various physically pertinent regimes. This determines the strengths and weaknesses of each analytical model, from the static Langevin function derived from Boltzmann statistics \cite{AM}, to the weak field Debye model \cite{Debye,Shl}, and lastly the low frequency effective field model \cite{Feldy}.

\section{Other approaches to modeling MNPs}

Currently there is no closed form solution to the stochastic Langevin equation, yet insight can often be teased from the rotational diffusion equation, sometimes also referred to as the Fokker-Planck equation. This approach introduces a distribution function for nanoparticle magnetization that carries all the associated probability moments. In contrast, the stochastic method has a more transparent differential equation, but requires solving for all the moments separately through repeated trials. As in Felderhof and Jones' paper \cite{Feldy} the orientation of a nanoparticle is presumed to satisfy the Einstein-Smoluchowski equation for the distribution function $f(\mathbf{m},t)$, where the azimuthal symmetry of the dipole potential energy $U=-\mathbf{m} \cdot \mathbf{B}$ results in a simplified relation that is only a function of the polar angle $f(\theta,t)$,
\begin{equation}\frac{\partial f}{\partial t}=D \left[\frac{1}{\sin \theta} \frac{\partial}{\partial \theta}\left(\sin \theta \frac{\partial f}{\partial \theta} + E(t) \sin^2 \theta f\right) \right]. \label{FkP} \end{equation}
The diffusion constant $D$ is the same as the stochastic approach above. The variable $E(t) =  mB\sin{(\omega t)}/k_B T$ is a ratio of magnetic to thermal energy with a perfect sinusoidal field. Eq.~\ref{FkP} could be used to examine the exact distribution function over magnetic moment angles but has not admitted an analytical solution. 

In 1929, Peter Debye developed a first order approximation for small driving fields that neglects non-linear behavior, and for our purposes is particularly limited to cases with no magnetic saturation \cite{Debye}. Linear response theory as used in Debye's method leads to a susceptibility that is a combination of real (in phase) and complex (out of phase) components: 
\begin{equation}\chi = \chi'+i\chi''\end{equation}
so that the magnetization parallel to the driving field can be expressed in terms of the Brownian relaxation time (Eq.~\ref{TB}),
an approximation in and of itself combining multiple relaxation mechanisms into one time constant. The Debye magnetization along the direction of the applied field is thus expressed
\begin{equation} M(t) = \frac{M_0}{1+(\omega\tau_B)^2} \left(\cos\omega t +\omega \tau_B \sin\omega t \right). \label{debEQ}\end{equation}
Both perturbative methods \cite{MW} and a series expansion in Legendre polynomials \cite{Raikher} have been used to develop a more complete solution to Eq.~\ref{FkP}. A particularly useful result, the so called effective field model, assumes that the frequency of oscillation is low enough so that the equilibrium distribution remains Maxwellian $f_{eq} \propto \exp({-U/k_BT})$. In low frequency regimes the quasi-static formulation does produce accurate nonlinear dynamics for the MNP magnetization with
\begin{equation}\frac{dM(t)}{dt}=-\frac{2M(t)}{\tau_B}\left( 1-\frac{E(t)}{\alpha_e(t)} \right) \label{effEQ} \end{equation}
where again $\tau_B$ is the Brownian relaxation time (Eq.~\ref{TB}), and at any moment in time, $\alpha_{e}(t)=m B/k_BT$ is the argument in the Langevin function,
\begin{equation} L(\alpha) = \coth \alpha - \frac{1}{\alpha}, \label{langEQ} \end{equation}
changing in time but not necessarily equal to the perfectly sinusoidal variation $E(t)$ from Eq.~\ref{FkP}. This slight difference encodes the non-linearity into the magnetization to first approximation. In practice the model is used to find the magnetization in the direction of the applied field by iterating the differential equation (Eq.~\ref{effEQ}) while at each time step inverting the Langevin function to find the proper effective field $\alpha_e$. 

Aside from the two dynamical models we've seen, the Langevin function itself provides a benchmark for the magnetization $M_{eq} = M_0\hspace{1mm}L(\alpha)$ in the static field case, provided the system is allowed to reach equilibrium. The three models presented will be compared to our stochastic model later in the work and the strengths and weaknesses of each will be discussed.
\section{Numerical methods}

To simulate the particle dynamics, we developed a Monte-Carlo scheme for MNPs in various fields and environmental conditions. To do so, the numerical recipe below (Eq.~\ref{recipe}), was implemented to solve the stochastic differential equation (Eq.~\ref{sde}) iteratively. Because at root we must admit that we are approximating some colored noise process as a white noise process, we take recourse to the Stratanovich interpretation of stochastic integration. The transformation also has practical benefits because the Stratanovich calculus preserves the rules of ordinary differential calculus (i.e. the chain rule) \cite{oks}. A fictitious drift term arises during the transformation from the It$\bar{\mathrm{o}}$ calculus and hence we can employ a stochastic numerical integration scheme which is transparently analogous to common methods. The simplest option in the stochastic realm, the Euler-Marayuma scheme (see for example Ref.~\cite{Gard} Ch.7) is sufficient because the slightly increased accuracy achieved through higher order methods is outweighed by longer computation times. We interpret the white noise force accordingly as a Wiener process $\mathrm{d}\mathbf{W}$, the derivative of a Markov process and also uncorrelated in varying space or time. It can be approximated numerically as a Gaussian distribution scaled by $\sqrt{\Delta t}$ \cite{Gard}. This results in the numerical algorithm
\begin{eqnarray*} \mathbf{m} + \Delta\mathbf{m}&=& \frac{m}{6\eta V}\left( \mathbf{m} \times \mathbf{H} \times \mathbf{m} - 2k_BT\mathbf{m} \right) \Delta t  \\ && +\sqrt{\frac{2k_BT}{6\eta V}\Delta t} \hspace{1mm} \mathbf{W} \times  \mathbf{m} \label{recipe}\end{eqnarray*}
where $m$ is the magnetic moment magnitude and $\mathbf{W}$ is a vector of random numbers with zero mean and unit standard deviation in each cartesian direction. At this level of analysis, single particle trajectories are practically meaningless. The quantities of interest are instead statistical moments, often requiring thousands to millions of samples. To examine equilibrium magnetization, simulation lengths were multiples of theoretical time constants. For simulations using oscillating fields, equilibrium dynamics were determined by converging similarity between periods. The standard deviation between a cycle and its predecessor divided by the amplitude of oscillation was used a metric for convergence. Cycles were repeated until this value was less than the threshold of 1\%. The constant numerical parameters were chosen and varied around those typical of MSB experiments. The mass magnetization was taken from a nanoparticle supplier (Micromod) data sheet as 76 emu/g for 100nm diameter particles. The particles are actually made up of many crystal domains, but are treated as though the bulk properties mimic the component properties and density is assumed constant so that the total magnetic moment can be written in its more common units of Joule/Tesla. The viscosity of the solution is assumed to be $\sim$1cP, near that of water at room temperature (293K). We pick realistic properties in order to see the correct physics in each regime of interest and this aids in the transition to quantitative comparison with experiments. 

\section{Model benchmarking}

\begin{figure}[h!]
\begin{center}
\includegraphics[width=2.8in]{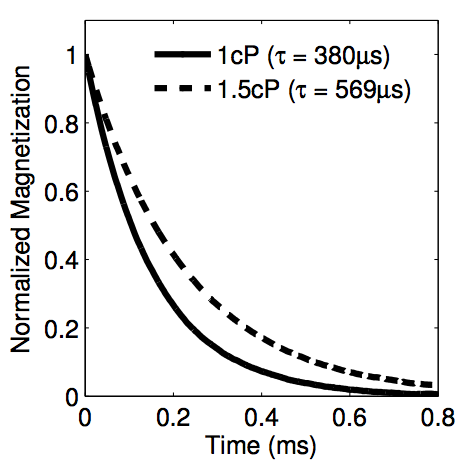}
\caption{Two relaxation curves with associated relaxation times found from exponential fits of the stochastic data. The expected relaxation times are given by Einstein's equation for Brownian relaxation to be 379.42$\mu$s and 569.13$\mu$s respectively, within nanoseconds of the simulated data. The simulations used Eq.~\ref{recipe} at two viscosities. $10^6$ particles were averaged beginning all in the z-direction, with no magnetic field.}
\label{visc}
\end{center}
\end{figure}
In this section, we verify that the model predicts well known physical results including a) the static field Langevin function, b) the high-frequency, low-amplitude Debye model, and c) the effective field model. The first benchmark is that with no applied field, the relaxation to equilibrium is on a time scale determined by Einstein's formulation of Brownian motion (Eq.~\ref{TB}). If a sample of particles is prepared to align in some direction but experience no external force, their average magnetization quickly decays to zero. By fitting to the normalized data an exponential decay of the form $M = e^{-t/\tau}$, we obtain the relaxation time. Increasing the viscosity intuitively increases the time constant. Fig.~\ref{visc} shows the stochastic model agrees with Einstein's theory with nanosecond precision. It should be noted however that as the viscosity is decreased, and particularly at the limit where $\eta \rightarrow 0$, Einstein's model becomes unphysical and the stochastic model predicts longer relaxation times.

\begin{figure}[h!]
\begin{center}
\includegraphics[width=2in]{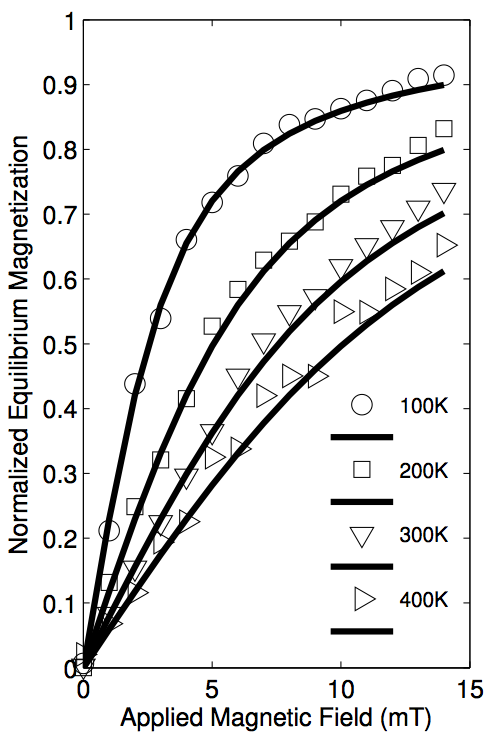}
\caption{The stochastic model (Eq.~\ref{recipe}, data points) is compared to the Langevin function (Eq.~\ref{langEQ}, solid lines) at various temperatures. The particles are initially aligned in the x-direction and the field points in the z-direction. $10^4$ particles were averaged, few enough to demonstrate that as temperature increases, the disorder in the system becomes much more prevalent.}
\label{lang}
\end{center}
\end{figure}

\begin{figure}[h!]
\begin{center}
\includegraphics[width=2.5in]{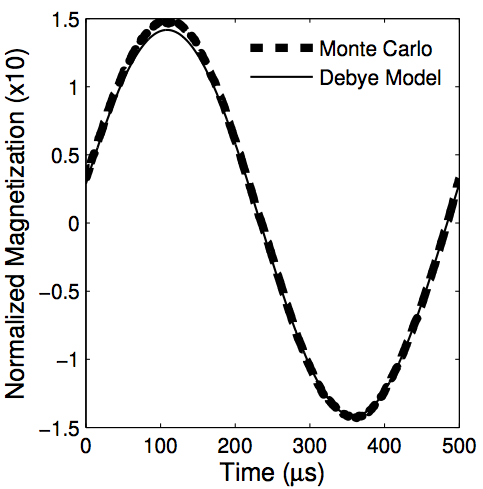}
\caption{In the low-field, high-frequency regime, the stochastic model (Eq.~\ref{recipe}) agrees with the Debye model (Eq.~\ref{debEQ}). Shown is the 20th cycle using $10^5$ particles in a 20mT, 2MHz field. In this simulation, the driving field is a cosine so the phase lag can be seen. The almost complete out of phase nature of the curve can be understood by examining the Debye model. Given a relaxation time of hundreds of microseconds and a frequency of 20kHz, the out of phase component is an order of magnitude larger than the in phase signal.} 
\label{deb}
\end{center}
\end{figure}

To test the model's behavior in a static field we begin with a sample of particles initially in the x-direction, and  employ a static field in the z-direction. The magnetization is allowed to evolve, and the value after five time constants is recorded. This gives a good estimate of the equilibrium magnetization and agrees with Langevin's theory as expected. Raising the viscosity does not change the outcome of the magnetization value, only delaying or expediting equilibrium. Increasing the temperature allows for more thermal motion, and therefore lessens the total alignment with the field effectively decreasing the final magnetization value. The data are depicted in Fig.~\ref{lang}.
Many models of MNPs benchmark against the equilibrium and static case, but we have extended ours to oscillating fields in order to corroborate analytic approximations. Here we begin by considering an oscillating field of the form $B = B_0\sin{\omega t}$ applied with high frequency ($\omega/2\pi =$ 20kHz). In this case, the magnetization does not saturate and follows the behavior of the Debye Model (see Fig.~\ref{deb}). 

\begin{figure}[h!]
\begin{center}
\includegraphics[width=3.375in]{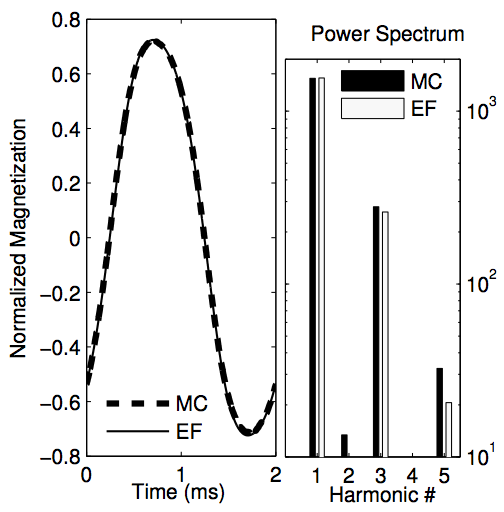}
\caption{The Monte-Carlo code (Eq.~\ref{recipe}) matches the effective field model magnetization (Eq.~\ref{effEQ}) and harmonic structure with a 20mT applied field of frequency 1kHz. The temperature is 300K and $10^5$ averages are used for a smooth curve. Note the applied field is sinusoidal and the phase-lag is evident.}
\label{effF}
\end{center}
\end{figure}

However, in a regime where particles come closer to saturation, the magnetization obeys different dynamics, including much more severe hysteretic effects, so that a simple linearization method like Debye's can no longer be applied. Described previously, the effective field model (Eq.~\ref{effEQ}) uses an adaptive method to solve for the effective thermal field from the inverse Langevin function and includes relaxation effects. Simulations demonstrate equivalence (Fig.~\ref{effF}) to the effective field model. Yet, the stochastic model extends past the range of the effective field model, which is only valid for low frequencies. Furthermore, our model replicates the correct behavior of solely odd harmonics from symmetric distortion. This is an important result because magnetization harmonics are used exclusively in practical imaging and sensing \cite{GW,rauw}.

\section{Experimental comparison and validation}

Magnetic particle spectroscopy can be used to discover information about the environment surrounding nanoparticles. One of the first experiments to do this employed MNPs as microscopic thermometers \cite{tempest}. To do so, a scaling relationship between the temperature and the magnetic field in the argument of the Langevin function was exploited. The method makes use of the low frequency, adiabatic limit assumption -- that the system is assumed to always be in equilibrium and thus governed by the Langevin function. Then if a control curve is developed for a known temperature, the scaling factor necessary to map subsequent curves onto the control can be used to quantify temperature. Note that ratios of higher harmonics are used for experimental reasons; to avoid dependence on particle number or systematic receive coil errors. Because the simulations have validated the Langevin function explicitly, we should not be too surprised to find that the scaling of field and temperature in the Langevin argument is consistent when attempted at low frequencies. Yet, that it works for the harmonic ratios (see Fig.~\ref{TB}) is an important confirmation. In the simulations, we have neglected any temperature dependence on viscosity or magnetic moment, and the adiabatic approximation fails above a few hundred cycles per second, at which point the harmonics do not carry enough information to distinguish between temperatures.

\begin{figure}[h!]
\begin{center}
\includegraphics[width=3in]{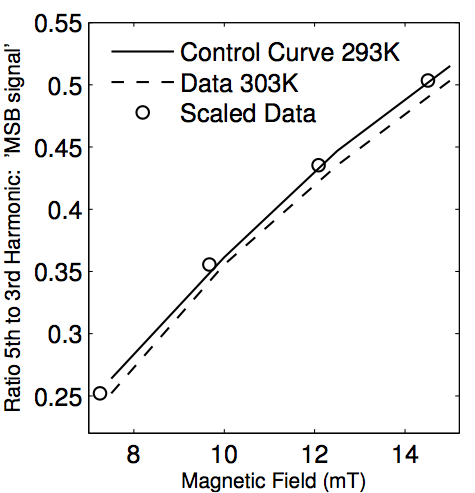}
\caption{A scaling argument can be made for temperature estimation of nanoparticles as in \cite{tempest}. By sweeping through field strengths and calculating the harmonic ratio for different temperatures, we show that scaling the x-values by the ratio of the control to new temperature shifts the data back on top of one another. Thus for an unknown temperature curve, a least squares fit can determine the scaling constant and the temperature can be recovered. In this figure, a field of 200Hz was applied at typical experimental values of 7.5-15mT. The two curves used $10^4$ particles at known temperatures of 293K and 303K respectively so in this case the scaling factor is 0.96.}
\label{TB}
\end{center}
\end{figure}

\begin{figure}[h!]
\begin{center}
\includegraphics[width=3in]{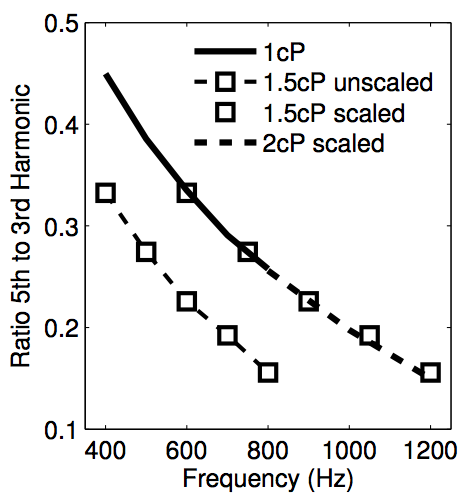}
\caption{A depiction of three frequency sweeps taking the harmonic ratio of the fifth to third harmonic as in \cite{jbwk}. The higher viscosity appears lower than the original, but can be scaled horizontally (in frequency) to align with the old curve. This compensates for viscous affects that are characterized by the relaxation time. The 1.5cP curve is shown explicitly before and after shift, while the 2cP is only showed post scaling, to justify that the three curves all do align. The simulations used a 20mT, 1kHz field, 300K, and four runs of $10^4$ particles each.}
\label{scale}
\end{center}
\end{figure}

Results also agree with experiments that determine nanoparticle relaxation time due to viscosity shifts \cite{jbwk}. Taking ratios of harmonics at various frequencies, there is a characteristic shift in time which compensates for the increasing relaxation time from an increased viscosity. By scaling the frequency range by the appropriate amount, the data can be shifted back onto a control curve. Then by measuring the relaxation time through the scaling value, obtained with a least-squares regression, the nanoparticles can be used as viscometer. The simulations are able to repeat the experimental method (Fig.~\ref{scale}). When scaled by viscosity, the curves align. Furthermore, by changing the specific values of the nanoparticles size and magnetic moment slightly from the values given by the manufacturer, the actual experimental data from Rauwerdink's paper can be reproduced.

\section{Conclusions}

A Langevin-type stochastic differential equation for the magnetization of non-interacting isotropic MNPs was developed and a numerical integration scheme was applied to model the particle dynamics in various types of magnetic fields. The simulations successfully modeled the approach to equilibrium (through the classical Einstein relaxation time) and the eventual state of thermal equilibrium was validated by the Langevin function at various temperatures. In oscillating fields, our approach agreed with other standard analytical approximations, surpassing the ability of both by achieving accuracy through a wide frequency range. The agreement included the magnetization shape as well as it's harmonic content in a distorted regime. Finally, it was shown that the model reproduces two experimental results from previous work on MSB sensing: that the relaxation time shift due to viscosity changes can be compensated for by a scaling in frequency so that nanoparticle solution viscosity can be measured, and that a similar method can be used in the adiabatic limit to measure temperature by scaling field strength. One of the large benefits of using stochastic analysis is the transparency of the differential equation. In the future, we will investigate further physics including the potential of decoupling temperature and viscosity effects into separate relaxation times.

\section{Acknowledgements}

This work was supported by an NIH-NCI grant 1U54CA151662-01.

\end{document}